\documentclass[aps,prx,twocolumn,superscriptaddress,longbibliography]{revtex4-2}

\usepackage{color}
\usepackage{soul}
\usepackage{dcolumn}
\usepackage{graphicx}
\usepackage{amsmath}
\usepackage{amssymb}
\usepackage{bm}
\usepackage[hidelinks]{hyperref}
\usepackage{multirow}

\begin{document}

\preprint{APS/123-QED}

\title{Fast logic with slow qubits:\\ microwave-activated controlled-Z gate on low-frequency fluxoniums}

\author{Quentin Ficheux}
\thanks{Equal contribution author}
\affiliation{Department of Physics, Joint Quantum Institute, and Center for Nanophysics and Advanced Materials, University of Maryland, College Park, MD 20742, USA}

\author{Long B. Nguyen}
\thanks{Equal contribution author}
\affiliation{Department of Physics, Joint Quantum Institute, and Center for Nanophysics and Advanced Materials, University of Maryland, College Park, MD 20742, USA}

\author{Aaron Somoroff}
\affiliation{Department of Physics, Joint Quantum Institute, and Center for Nanophysics and Advanced Materials, University of Maryland, College Park, MD 20742, USA}

\author{Haonan Xiong}
\affiliation{Department of Physics, Joint Quantum Institute, and Center for Nanophysics and Advanced Materials, University of Maryland, College Park, MD 20742, USA}

\author{Konstantin N. Nesterov}
\affiliation{Department of Physics and  Wisconsin Quantum Institute, University of Wisconsin - Madison, Madison, WI 53706, USA}

\author{Maxim G. Vavilov}
\affiliation{Department of Physics and  Wisconsin Quantum Institute, University of Wisconsin - Madison, Madison, WI 53706, USA}

\author{Vladimir E. Manucharyan}
\affiliation{Department of Physics, Joint Quantum Institute, and Center for Nanophysics and Advanced Materials, University of Maryland, College Park, MD 20742, USA}

\date{\today}

\begin{abstract}
We demonstrate a controlled-Z gate between capacitively coupled fluxonium qubits with transition frequencies $72.3~\textrm{MHz}$ and $136.3~\textrm{MHz}$. The gate is activated by a $61.6~\textrm{ns}$ long pulse at the frequency between non-computational transitions $|10\rangle - |20\rangle$ and $|11\rangle - |21\rangle$, during which the qubits complete only $4$ and $8$ Larmor periods, respectively. The measured gate error of $(8\pm1)\times 10^{-3}$ is limited by decoherence in the non-computational subspace, which will likely improve in the next generation devices. Although our qubits are about fifty times slower than transmons, the two-qubit gate is faster than microwave-activated gates on transmons, and the gate error is on par with the lowest reported. Architectural advantages of low-frequency fluxoniums include long qubit coherence time, weak hybridization in the computational subspace, suppressed residual $ZZ$-coupling rate (here $46~\mathrm{kHz}$), and absence of either excessive parameter matching or complex pulse shaping requirements.\\

\end{abstract}

\maketitle

\section{Introduction}

Macroscopic superconducting circuits have emerged as a leading platform for implementing a quantum computer~\cite{Kjaergaard2020a}. Currently available small-scale superconducting quantum processors \cite{Kelly2015,Otterbach2017,Kandala2017,Song2017a,Neill2018,Wei2020,Jurcevic2020,Hashim2020} have achieved a number of important milestones, including the break-even point in quantum error correction of a single logical qubit \cite{Ofek2016a}, digital quantum simulation \cite{Heras2014,Salathe2015,Barends2016,Arute2020,Fedorov2020,Fauseweh2020}, non-trivial optimization algorithms \cite{Barends2015,Moll2018,Arute2020a}, and, an example demonstration of quantum supremacy with 53 qubits \cite{Arute2019}. This progress is even more spectacular as it is solely based on a single qubit type, the transmon, which is essentially a weakly-anharmonic electromagnetic oscillator \cite{Koch2007}. Although the transmon's simplicity makes it an exceptionally robust quantum system, its relatively weak non-linearity has become a major limiting factor for current performance and further scaling of quantum processors. Irrespective of implementation details, a weak anharmonicity inevitably leads to slower two-qubit gates, which makes them prone to decoherence errors. Therefore, a motivation has built up for exploring strongly-anharmonic alternatives to transmons that would ideally have higher intrinsic coherence and be compatible with the transmon-based scaling architectures.

In recent years, long coherence times (up to $T_2 =500~\mu s$) were repeatedly observed in superconducting fluxonium qubits \cite{Nguyen2018,Zhang2020}. In this paper we describe the first logical operation on a pair of capacitively-coupled fluxoniums [Fig.~1(a)]. 
Each fluxonium has one weak Josephson junction and a large capacitor, which are the same components as in the transmon, but the junction is additionally shunted by a large-value inductance \cite{Manucharyan2009}, made from a chain of about 100 stronger junctions. Inductive shunting makes the charges on the fluxonium capacitors continuous, and hence, contrary to transmons, fluxoniums can have highly anharmonic spectra insensitive to offset charges. We operate fluxoniums near the half-integer flux bias (``sweet spot"), where the qubit transition frequency is first order insensitive to flux noise and belongs to a $100-1000~\textrm{MHz}$ range. The order of magnitude qubit slowdown, compared to typically $5-6~\textrm{GHz}$ transmons, dramatically reduces the rate of energy relaxation due to dielectric loss, an important decoherence mechanism in superconducting circuits. This straightforward trick proved largely responsible for the long coherence time of fluxoniums \cite{Nguyen2018}. However, how can one slow qubits down without ending up with slower two-qubit gates?

To answer this general question, we consider the simplest form of circuit-circuit coupling by means of a mutual capacitance. In analogy with transmons, we get a coupling term proportional to $n_A n_B$, where $n_{A}$ and $n_{B}$ are the charge operators of qubits A and B, normalized to the Cooper pair charge $2e$. First, we notice that capacitive coupling indeed produces little effect on the computational states $|00\rangle$, $|01\rangle$, $|10\rangle$, $|11\rangle$, because the transition matrix elements of $n_{A(B)}$ vanish with the transition frequency. For the same reason, the much higher energy non-computational states $|12\rangle$ and $|21\rangle$ can experience a noticeable repulsion, while the only nearby states $|20\rangle$ and $|02\rangle$ remain unaffected due to the parity selection rule, see Fig.~1(b). Therefore, connecting the two subspaces with radiation can induce an on-demand qubit-qubit interaction. For example, Ref. \cite{Nesterov2018} describes a controlled-Z (CZ) gate obtained by applying a $2\pi$-pulse to transition $|11\rangle - |21\rangle$ while the closest transition $|10\rangle - |20\rangle$ stays unexcited. This condition can always be met if the gate pulse is much longer than $1/\Delta$, where $\Delta$ is the shift of level $|21\rangle$ due to the $n_An_B$-term, shown in Fig.~1(b). Yet, we show that with an optimal combination of drive detuning and amplitude, the CZ-gate can be completed in a time close to $1/\Delta$ with zero leakage outside the computational subspace, thanks to synchronization of Rabi rotations of both non-computational transitions. For our specific device parameters (see Tables \ref{table:device_parameters}, \ref{table:T1T2}), we get $\Delta = 22~\textrm{MHz}$ ($1/\Delta = 45.5~\textrm{ns}$) and the optimal gate time is $61.6~\textrm{ns}$. Remarkably, since $\Delta$ is not tied to the qubit frequencies (here $72.3~\textrm{MHz}$ and $136.3~\textrm{MHz}$), the logical operation takes just a few qubit Larmor periods. To our knowledge, such a high relative gate speed is unmatched across quantum computing platforms. 

Population transit through non-computational states is common in gates realized with transmons \cite{Strauch2003,Dicarlo2009, Abdumalikov2013, Egger2014, Barends2019, Krinner2020,Ganzhorn2020, Negirneac2020}. For example, repulsion of states $|11\rangle$ and $|20\rangle$ enables a CZ-gate via flux-tuning of these states in and out of resonance \cite{Dicarlo2009}. Recent high-fidelity versions of this gate rely on diabatic flux pulses \cite{Barends2019,Negirneac2020}, resulting in a significant population of state $|20\rangle$ for a short time, which draws parallels to our microwave-controlled scheme. In the case of fixed-frequency qubits, repulsion between states $|30\rangle$ and $|21\rangle$ was used in Ref.~\cite{Chow2013a} to implement a CZ-gate activated by a microwave pulse at a frequency near the transition $|11\rangle - |12\rangle$. However, the insufficient transmon anharmonicity introduces many nearby transitions (there is only one relevant transition $|10\rangle - |20\rangle$ for fluxoniums), and in the end such a gate proved impractical. Even in gates designed to operate entirely within the computational subspace, e.g. the flux-activated $|10\rangle - |01\rangle$ swap gate \cite{Blais2003,Bialczak2010,Dewes2012} or the cross-resonance gate \cite{Rigetti2010,Chow2011}, uncontrolled population leakage to non-computational states remains an important factor limiting gate speed \cite{Sheldon2016}. Yet, such coherent errors can be practically eliminated in fluxoniums owing to their highly anharmonic spectra, as exemplified by the gate scheme reported here.

\begin{figure}[t]
    \centering
    \includegraphics[width=\columnwidth]{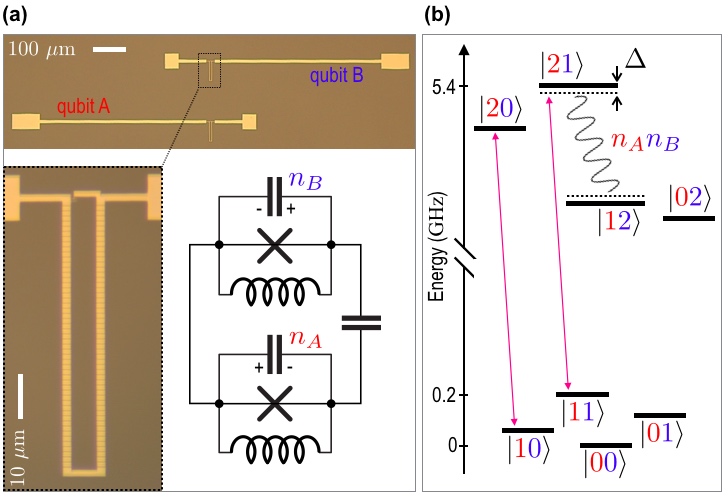}
    \caption{
    (a) Optical image of two capacitively-coupled fluxoniums fabricated on a silicon chip along with their minimal circuit model. Devices are similar to those reported in Ref. \cite{Nguyen2018} except the antennas are intentionally made asymmetric for optimal coupling to the readout cavity [not shown]. (b) Diagram of the lowest energy states of the interacting two-fluxonium system. Capacitive coupling induces a shift of level $|21\rangle$ by $\Delta$ due to repulsion from level $|12\rangle$. The shift $\Delta$ leads to a CZ-operation in time approximately given by $1/\Delta$ when either qubit is driven at a frequency in between the transitions $|10\rangle - |20\rangle$ and $|11\rangle - |21\rangle$.
    }
    \label{fig:fig1}
\end{figure}

Another important advantage of fuxoniums over fixed-coupling transmons is that the static $ZZ$-shift, coming from the repulsion of computational and non-computational states, can be suppressed by over an order of magnitude compared to typical values in capacitively coupled transmons (here about $40~\textrm{kHz}$), largely thanks to the low qubit frequencies. For transmons, the static $ZZ$-shift is an important source of gate error, the mitigation of which draws additional resources. Thus, in the case of the cross-resonance gate, the $ZZ$-term is suppressed by a combination of circuit parameter matching, additional echo pulse sequences incorporated into the gate protocol \cite{Sheldon2016}, and additional qubit rotations \cite{Sundaresan2020}. An alternative strategy to eliminate the $ZZ$-shift is to use flux-tunable couplers \cite{Mundada2019,Arute2019,Li2019}, which in practice act as separate quantum systems and hence increase the circuit complexity. More recently, the $ZZ$-shift was suppressed in capacitively-shunted flux qubits using an additional drive, but at the expense of operating away from the flux sweet-spot \cite{Noguchi2020}.

The gate error in our scheme is largely limited by decoherence outside the computational subspace. Randomized benchmarking yields the averaged gate error of $8\times 10^{-3}$, which is consistent with the measured few microseconds coherence time of transitions $|10\rangle - |20\rangle$ and $|11\rangle - |21\rangle$ (see Table \ref{table:T1T2}). Because these transitions are transmon-like, we expect their coherence will improve by an order of magnitude in the next generation experiments with improved fabrication and noise filtering procedures. This step would lower the gate error into the $10^{-4}$-range. The presently achieved fidelity is on par with the lowest reported values in microwave-activated gates \cite{McKay2019,Sundaresan2020} (cross-resonance gate by IBM). Additionally, our gate is considerably faster than  cross-resonance type gates with comparable errors \cite{Corcoles2013,Sheldon2016}. Our result illustrates the potential of highly-anharmonic circuits for quantum information processing and it motivates the exploration of large-scale quantum processors based on fluxoniums.

The paper is organized as follows, in Sec. II we describe the details of our experimental setup, including spectroscopy of the two-fluxonium device, the joint single-shot readout of fluxoniums, and state initialization procedures. In Sec. III we detail the concepts behind our fast microwave-activated CZ-gate. Section IV presents the results of gate characterization, including quantum process tomography and randomized benchmarking. In Sec. V we review the technical limitations of the present experiment, and project the near term improvements. Section VI concludes the work. 

\section{\label{sec:characterization}Two-fluxonium system characterization}

Our device is composed of two fluxonium artificial atoms with a circuit design introduced in \cite{Nguyen2018} coupled via a shared capacitance [see Fig.~\ref{fig:fig1}(a)]. The system obeys the Hamiltonian \cite{Vool2017, Nesterov2018}
\begin{equation}
\begin{aligned}
\frac{\hat{H}}{h} =  & \sum_{\alpha = A,B} 4 E_{C,\alpha} \hat{n}_\alpha^2 + \frac{1}{2}E_{L,\alpha} \hat{\varphi}_\alpha^2 - E_{J,\alpha} \cos(\hat{\varphi}_\alpha-\phi_{\mathrm{ext},\alpha})  \\ &+ J_C \hat{n}_A \hat{n}_B
\label{eq:Hamiltonian}
\end{aligned}
\end{equation} where $E_{C,\alpha}$, $E_{L,\alpha}$, and $E_{J,\alpha}$ are the charging energy, the inductive energy and the Josephson energy of fluxonium indexed by $\alpha = A, B$, respectively. The operators $\hat{\varphi}_\alpha$ and $\hat{n}_\alpha$ are the phase twist across the inductance $L_\alpha$ and the charge on the capacitor $C_\alpha$, and they commute according to $[\hat \varphi_\alpha,\hat n_\alpha] = i$. The experimentally extracted parameters of the Hamiltonian (1) are given in Table~\ref{table:device_parameters}.

\begin{center}
\begin{table}[h!]
\begin{tabular}{ | c | c | c | c | c |} \hline
  & $E_{C,\alpha} \mathrm{~(GHz)}$ & $E_{L,\alpha} \mathrm{~(GHz)}$ & $E_{J,\alpha} \mathrm{~(GHz)}$ & $J_C \mathrm{~(GHz)}$\\ \hline
  Qubit A & $0.973$ & $0.457$ & $5.899$&  \multirow{2}{*}{0.224}
 \\ 
  Qubit B & $1.027$ & $0.684$ & $5.768$& \\ \hline
\end{tabular}
\caption{\label{table:device_parameters} Parameters of Hamiltonian Eq.~(\ref{eq:Hamiltonian}) extracted by its fitting to the two-tone spectroscopy data in Fig.~\ref{fig:fig1b}.
}
\end{table}
\end{center}

\begin{figure}
    \centering
    \includegraphics[width=\columnwidth]{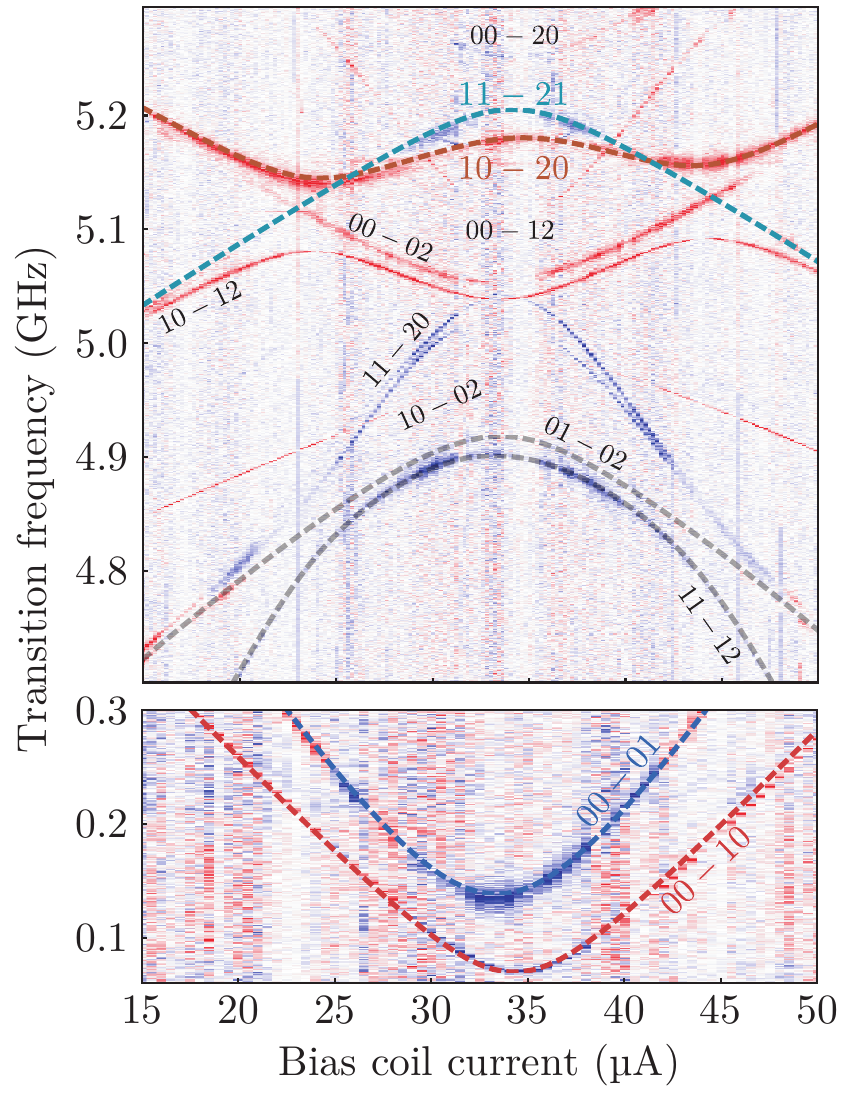}
    \caption{Two-tone spectroscopy as a function of probe tone frequency and flux threading the common biasing coil. The parameters of the system (Table \ref{table:device_parameters}) are extracted by fitting the spectroscopy data to the diagonalization of the Hamiltonian (\ref{eq:Hamiltonian}). Dashed lines represent transitions allowed at half-integer flux while dotted lines represent transitions that are forbidden at sweet spot. The shared capacitance lifts the degeneracy between the two $|11 \rangle - | 21 \rangle$ and $| 10 \rangle - | 20 \rangle$ transitions by an amount $\Delta$ (top pannel) that limits the gate speed. The sweet spots misalignment (bottom panel) is attributed to a local flux imbalance through the two fluxonium loops $\phi_{\mathrm{ext},A} \neq \phi_{\mathrm{ext},B}$.
    }
    \label{fig:fig1b}
\end{figure}

When biased at $\phi_{\mathrm{ext},\alpha} = \pi$, the fluxoniums are at their sweet spots with respect to external flux. The qubit frequencies $f_A = 72.3 \mathrm{~MHz}$ and $f_B= 136.3 \mathrm{~MHz}$ are almost two orders of magnitude lower than the typical transition frequency in cavity resonators and transmon qubits \cite{Schreier2008}. These low frequency transitions exhibit a long energy relaxation times $T_{1,A} = 347 \mathrm{~\mu s}$ and $T_{1,B} = 282 \mathrm{~\mu s}$ owing to their decoupling from dielectric loss mechanisms \cite{Nguyen2018}. A slight non-uniformity in the magnetic field, provided by a single external coil, prevents biasing qubits precisely at their sweet-spots simultaneously, with the offset being about $0.14\%$ of the flux quantum. We operate at a bias coil current of $33.7 \mathrm{~\mu A}$, approximately halfway between the two sweet spots, which reduced the spin echo coherence times to $T_{2,A}^E=31 \mathrm{~\mu s}$ and $T_{2,B}^E=64 \mathrm{~\mu s}$ (see Table \ref{table:T1T2}) compared to their sweet spot values of $47 \mathrm{~\mu s}$ and $67 \mathrm{~\mu s}$, respectively. The sweet-spot values of coherence times are likely limited by photon shot-noise due to imperfect thermalization of the readout cavity \cite{Schuster2005}. This common issue can be improved with better filtering of the measurement lines. We note in advance that the small deviation from the sweet-spots does not modify appreciably the operation of our two-qubit gates, and the qubit coherence times are not limiting the gate error.

\begin{figure*}
    \centering
    \includegraphics[width=0.95\textwidth]{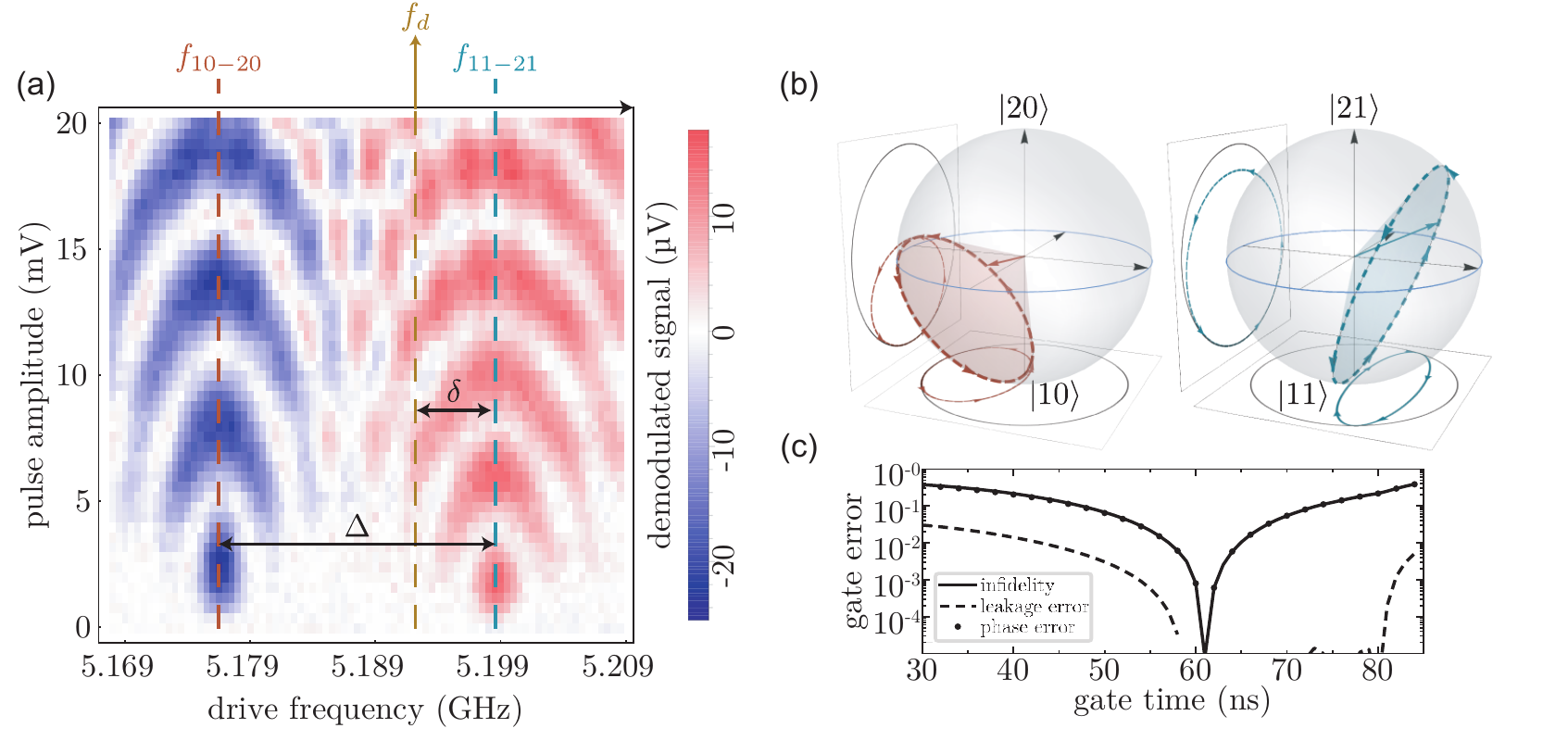}
    \caption{
    Gate principle. (a) Rabi oscillations near the $|10\rangle - |20 \rangle$ and $|11 \rangle - |21 \rangle$ transitions versus driving frequency with a $330 \mathrm{~ns}$-long pulse. The two transitions display different resonance  Rabi frequencies characterized by the ratio $r =\Omega_{11-21} / \Omega_{10-20}\simeq 1.36$. The drive frequency $f_d$ indicated in brown given by Eq.~(\ref{eq:drivefreq}) is used to synchronize the oscillations on the two transitions. (b) Bloch sphere representations of the trajectories in the $\{ |10 \rangle, |20 \rangle \}$ and $ \{ | 11\rangle, | 21 \rangle\}$ manifolds in the frame rotating at the drive frequency. The quantum state follows a closed path in the Hilbert space leading to a relative phase accumulation given by the difference of the solid angles spanned by the two trajectories. (c) Gate error versus gate duration simulated using the Hamiltonian given by Eq.~(\ref{eq:Hamiltonian}) in the presence of a Gaussian flat-topped pulse and without decoherence. For every gate duration, drive frequency and amplitude are optimized to minimize the coherent leakage error (dashed line). The dots represent the gate error caused by incorrect phase accumulation $\Delta \varphi = \varphi_{11}-\varphi_{10} \neq \pi$. With the optimal pulse duration around $\sim 62 \mathrm{~ns}$, the infidelity due to coherent errors is well below $10^{-3}$.
    }
    \label{fig:fig2}
\end{figure*}

Higher energy states are separated from the computational states by an approximately $4.5~\textrm{GHz}$ gap (Fig.~\ref{fig:fig1b} - top panel). The capacitive coupling term $J_c \hat{n}_A \hat{n}_B$ in Eq.~(\ref{eq:Hamiltonian}) with $J_C = 224 \mathrm{~MHz}$ manifests itself in the spectrum of lowest non-computational transitions. Mainly, the interaction creates a frequency splitting between transitions $|10 \rangle - |20 \rangle$ and $|11 \rangle - |21 \rangle$ of $\Delta = 22 \mathrm{~MHz}$.
Note that the extracted value of $J_C$ corresponds to the mutual capacitance between the qubit circuits of about $1.2\mathrm{~fF}$, which is comparable to the typical mutual capacitance used for coupling transmons in quantum processors. Owing to the matrix element hierarchy $n^\alpha_{0 - 1} \ll n^\alpha_{1 - 2}$, where $n^\alpha_{i - j}= |\langle i | \hat{n}_\alpha | j \rangle |$ is the charge matrix element of fluxonium $\alpha$, the transverse interaction in the computational subspace is negligible. Interaction involving higher levels also introduces a static longitudinal $ZZ$-interaction in the computational subspace \cite{Mundada2019,Noguchi2020} given by $\xi_{ZZ} = (E_{00}+E_{11}-E_{01}-E_{10})/h = 46 \mathrm{~kHz}$, where $E_{ij}$ is the energy of state $|i j \rangle$. We note that this value of $\xi_{ZZ}$ is at least an order of magnitude smaller than in typical transmon circuits and does not require extra spectrum optimizations beyond having low qubit-transitions frequencies.

The device is embedded in a 3-dimensional copper cavity with a resonant frequency of $f_c = 7.4806 \mathrm{~GHz}$. Prior to each experiment, the qubits are initialized by applying a strong microwave pulse on the cavity (see Appendix \ref{sec:initialization}). This procedure increases the excited state probability of the qubits to $86\%$ and $88\%$ which is sufficient for metrology of gate operations. The remaining qubit entropy corresponds to temperatures of $T_A = 3.7 \mathrm{~mK}$ and $T_B = 5.9 \mathrm{~mK}$ well below the fridge temperature of $14\mathrm{~mK}$. We achieve a single-shot joint readout of the two qubits by sending a cavity tone of optimal frequency, power, and duration. The outgoing signal is further amplified by a Josephson Traveling Wave Parametric Amplifier (JTWPA) \cite{Macklin2015} and commercial amplifiers before down-conversion and numerical demodulation. We finally correct for readout imperfections by a procedure described in Appendix \ref{sec:readoutcrosstalk}.

\section{\label{sec:concept} Fast CZ-gate by exact leakage cancellation}

In this section, we describe how to exploit the fluxonium interaction to implement a CZ gate in the shortest possible time. Let us start by considering the gate transitions $|10 \rangle - | 20 \rangle$ and $| 11 \rangle - | 21 \rangle$ as two-level systems. Our gate exploits the geometric-phase accumulations during  round trips in these two systems. The accumulated phase can be divided into two parts: the dynamical phase, which is proportional to the evolution time and the energy of the system, and the geometric phase, which depends only on the trajectory followed in the Hilbert space. Applying a microwave tone drives the system with the Hamiltonian
\begin{equation}
\frac{\hat{H}_\mathrm{drive}}{h} = (\epsilon_A \hat{n}_A + \epsilon_B \hat{n}_B) \cos( 2 \pi f_d t).
\label{eq:drive}
\end{equation} When the drive frequency $f_d$ is nearly resonant with the gate transitions, we observe Rabi oscillations [Fig.~\ref{fig:fig2}(a))] with the resonance Rabi frequencies $\Omega_{10-20} = | \langle 10 | \hat{H}_\mathrm{drive}| 20 \rangle | / h$ and $\Omega_{11-21} = | \langle 11 | \hat{H}_\mathrm{drive}| 21 \rangle | / h$, respectively. A strong hybridization of the $|12 \rangle$ and $|21 \rangle$ states creates an imbalance between the rotation speeds given by the ratio $r = \Omega_{11-21} / \Omega_{10-20}  \simeq 1.36$.

In order to eliminate leakage to higher states, one needs to synchronize off-resonance Rabi oscillations determined by the two transitions to ensure that the state vector always comes back to the computational subspace. This is achieved by matching the generalized Rabi frequencies 
\begin{equation}
\Omega = \sqrt{\Omega_{11 - 21}^2+\delta^2} = \sqrt{\Omega_{10 - 20}^2+(\delta-\Delta)^2} ,
\label{eq:syncrhonization}
\end{equation}
where $\delta = f_{11 - 21} - f_d$ is the detuning between the $| 11 \rangle - | 21 \rangle$ transition and the drive frequency [Fig.~\ref{fig:fig2}(a)]. A full rotation is then preformed in the shortest gate time $t_\mathrm{gate}= 1/\Omega $. During the gate operation, the state vector trajectory can be depicted in a Bloch sphere representation [Fig.~\ref{fig:fig2}(b)] when the system starts in $| 10 \rangle$ or $| 11 \rangle$. These circular trajectories, which are traveled in opposite directions with respect to the centers of the Bloch spheres, define two cones inside the spheres. The cones and directions of travel define the solid angles $\Theta_{10} = 2 \pi [1- (\Delta-\delta)/\Omega]$ and $\Theta_{11} = 2 \pi (1+\delta/\Omega)$, which correspond to a geometric phase accumulation $\varphi_{ij} = - \Theta_{ij}/2$ on state $|i j \rangle$. Our gate thus implements a unitary operation $U = \mathrm{diag}(1,1,e^{i \varphi_{10}}, e^{i \varphi_{11}})$. Using virtual Z rotations \cite{McKay2017}, the phase difference can be assigned to any computational state such as $\left| 11 \right>$ to realize a controlled-Phase operation $U=\mathrm{diag}(1,1,1,e^{i \Delta \varphi})$ with $\Delta \varphi = \varphi_{11}-\varphi_{10}$. A CZ gate is obtained when $\Delta \varphi = - (\Theta_{11}-\Theta_{10})/2= - \pi \Delta / \Omega= \pm \pi$. Using this condition in Eq.~(\ref{eq:syncrhonization}) we obtain the optimal drive frequency [brown arrow on Fig.~\ref{fig:fig2}(a)]
\begin{equation}
\frac{\delta}{\Delta} = \frac{r^2-\sqrt{(r^2-1)^2+r^2}}{r^2-1} \simeq 0.29\, .
\label{eq:drivefreq}
\end{equation}
For $\delta$ given by Eq.~(\ref{eq:drivefreq}) and $\Omega = \Delta$, a CZ gate with zero leakage is achieved in time exactly $t_\mathrm{gate} =1/\Delta$.

Our understanding of the gate process is validated by simulating the full Hamiltonian Eq.~\eqref{eq:Hamiltonian} in the presence of the drive Hamiltonian, Eq.~\eqref{eq:drive}.
The full Hamiltonian takes into account the dynamical phase $\phi_\mathrm{dyn} = 2 \pi \xi_{ZZ} t_\mathrm{gate} \simeq 10^{-2} \ll \pi$ which is negligible after one gate operation thanks to the small $ZZ$-interaction term. Starting with the driving frequency given by Eq.~\eqref{eq:drivefreq}, we minimize the final populations leakage out of the computational subspace by adjusting the drive amplitude and frequency for every gate duration. Because of the rising and lowering edges of the pulse in the simulation, we find a slightly longer optimal gate duration $\sim 61 \mathrm{~ns}$ (compared to $1/\Delta = 45.5~\textrm{ns}$) for which coherent errors on the gate fall below $<10^{-3}$, see Fig.~\ref{fig:fig2}(c). Note, even at the error level of $10^{-4}$, the optimal point does not require an excessive fine-tuning of the gate pulse parameters: it corresponds to a fraction of a percent variation in terms of the gate time, gate pulse amplitude, and frequency (see Appendix \ref{sec:driveparameters}). Our procedure can readily be extended to any other phase accumulation by imposing a different phase-accumulation condition leading to a different gate duration and modified single-qubit virtual Z rotations.

\section{\label{sec:calibration} Gate calibration and metrology}

\begin{figure}[t]
    \centering
    \includegraphics[width=0.95\columnwidth]{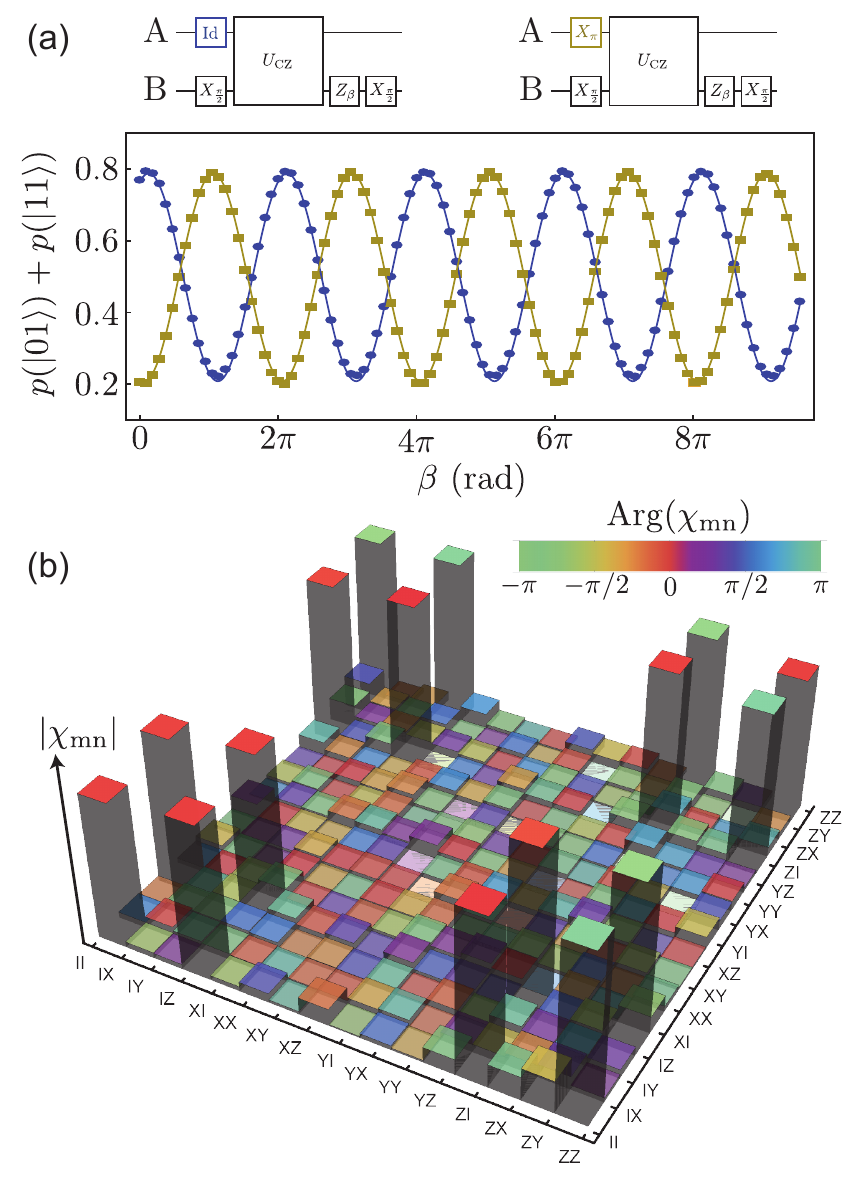}
    \caption{Controlled-Z gate. (a) Pulse sequences of a Ramsey-type experiment measuring the phase of qubit B after a CZ operation. Ramsey fringes are obtained by varying the rotation angle $\beta$. The excited state population of qubit B $p_1^B = p(|01 \rangle)+ p(|11\rangle)$ oscillates with $\beta$ with a phase that depends on the state of the control qubit A. The blue and brown fringes are obtained by flipping the state of qubit A at the beginning of the sequence revealing the conditional phase accumulated during the gate $\Delta \varphi \simeq 0.95 \pi$. (b) Quantum process tomography of the CZ gate. The experimentally extracted process tomography matrix $\chi$ reproduces a CZ operation with a fidelity of $0.97$.}
    \label{fig:fig5}
\end{figure}

Using our understanding from the previous section, we start with the following initial gate parameters: we use a $60 \mathrm{~ns}$ flat-top Gaussian pulse with a driving amplitude corresponding to a full rotation at the gate frequency given by Eq.~(\ref{eq:drivefreq}). To measure the accumulated phase difference $\Delta \varphi$, we perform a Ramsey-type experiment on qubit B to compare the phases of the superpositions $|11\rangle + |10 \rangle$ and $|01 \rangle + | 00 \rangle$ after the application of a CZ gate. This is achieved by inserting a CZ gate between two successive $\pi/2$ pulses with a relative phase $\beta$ created by a virtual Z rotation \cite{McKay2017}. When the angle $\beta$ is varied, we observe Ramsey fringes with an initial phase encoding the excited state probability of the control qubit A. The phase difference when qubit A is prepared in the ground or excited state yield the relative phase accumulation $\Delta \varphi / \pi \simeq 0.95$ [in Fig.~\ref{fig:fig5}(a)] close to the phase required for the CZ gate.

In order to gain more insight in the physics of our gate, we perform quantum process tomography by preparing $16$ independent input states, applying the CZ gate to each of them before performing their state tomography. The state tomography is obtained by a maximum likelihood estimation similar to the one in Ref. \cite{Roy2020} using an overcomplete set of $36$ tomography pulses. We represent the quantum process \cite{Nielsen2000} in the Pauli basis through the process tomography matrix $\chi$ \cite{Corcoles2013} [Fig.~\ref{fig:fig5}(b)] and adjust the single qubit phases with virtual Z rotations to attribute the relative phase accumulation to the $|11 \rangle$ state only. We obtain process fidelity $F_\mathrm{QPT}= 0.97$. This result shows that our scheme can be used to implement a CZ gate but the value of the fidelity is probably dominated by state preparation and measurement (SPAM) errors (Appendix \ref{sec:readoutcrosstalk} and \ref{sec:singlequbitgates}) and cannot be used to optimize the gate parameters further.

\begin{figure}
    \centering
    \includegraphics[width=\columnwidth]{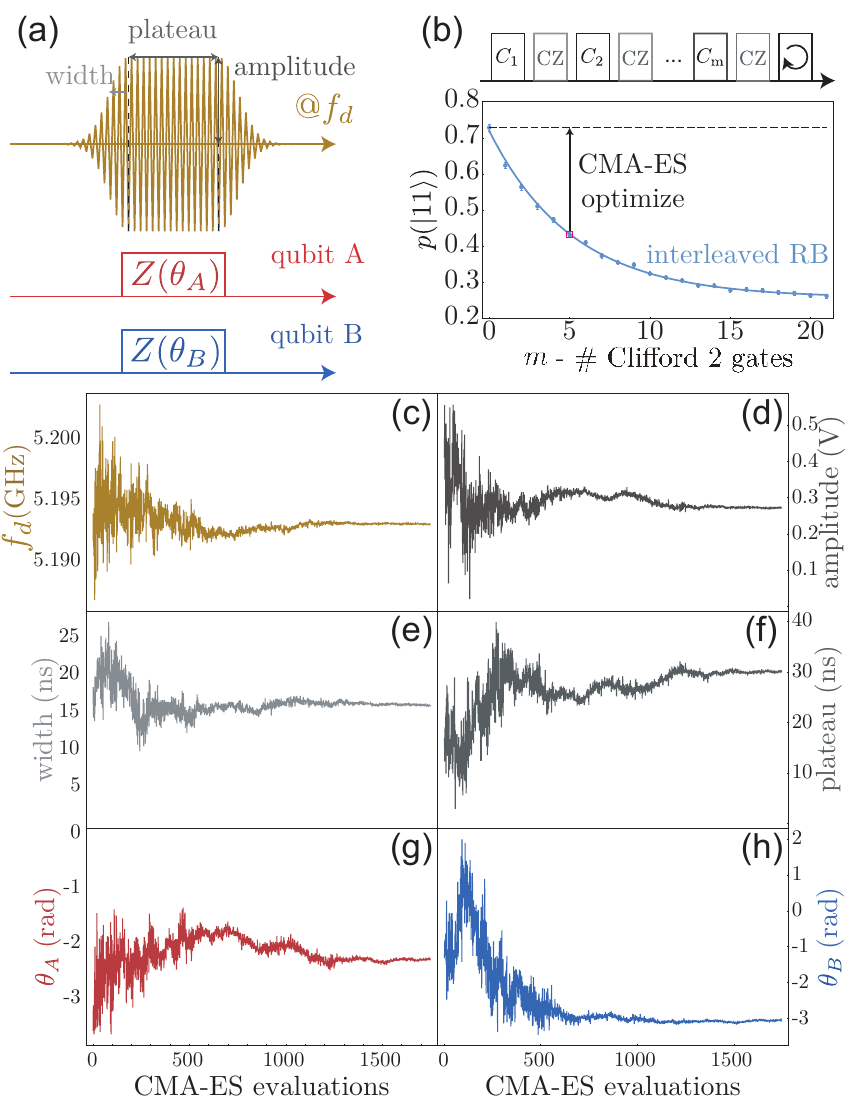}
    \caption{
    Optimizing the CZ gate. (a) The gate is optimized over $6$ parameters : pulse width, plateau, amplitude, and frequency as well as single qubit Z rotation angles. (b) We use an approach based on Optimized Randomized Benchmarking for Immediate Tune-up (ORBIT) to optimize the sequence fidelity at fixed length with a CMA-ES optimization. The number of Clifford gates [here $5 \simeq -1/\log(p)$] is chosen to maximize the sensitivity of the tune-up procedure. (c)-(h) Stochastic evolution of the gate parameters versus the number of function evaluation. The fidelity of the CZ gate is improved from $0.967$ to $0.992$.}
    \label{fig:fig3}
\end{figure}

Randomized benchmarking (RB) provides a reliable and robust metric of gate fidelity. The sequence is composed of $m$ randomly chosen Clifford operations followed by a recovery gate aimed at bringing back the quantum state to the initial state. The average population of $| 11 \rangle$ state decays as $a+b p^m+ c (m-1) p^{m-2}$ where $m$ is the number of Clifford operations, $p$ is the depolarizing parameter and $a,b,c$ are fitting parameters used to absorb state preparation and measurement errors \cite{Magesan2012a}. The average fidelity of a Clifford operation is given by $F_\mathrm{Clifford} = 1- \frac{d-1}{d}(1-p)$ where $d=2^n$ is the dimension of the Hilbert space with $n=2$ the number of qubits. Interleaving a target gate yields a decay with depolarizing parameter $p_\mathrm{gate}$ and a gate error of $1-F= \frac{d-1}{d}(1-p_\mathrm{gate}/p)$.

\begin{figure}[t]
    \centering
    \includegraphics[width=\columnwidth]{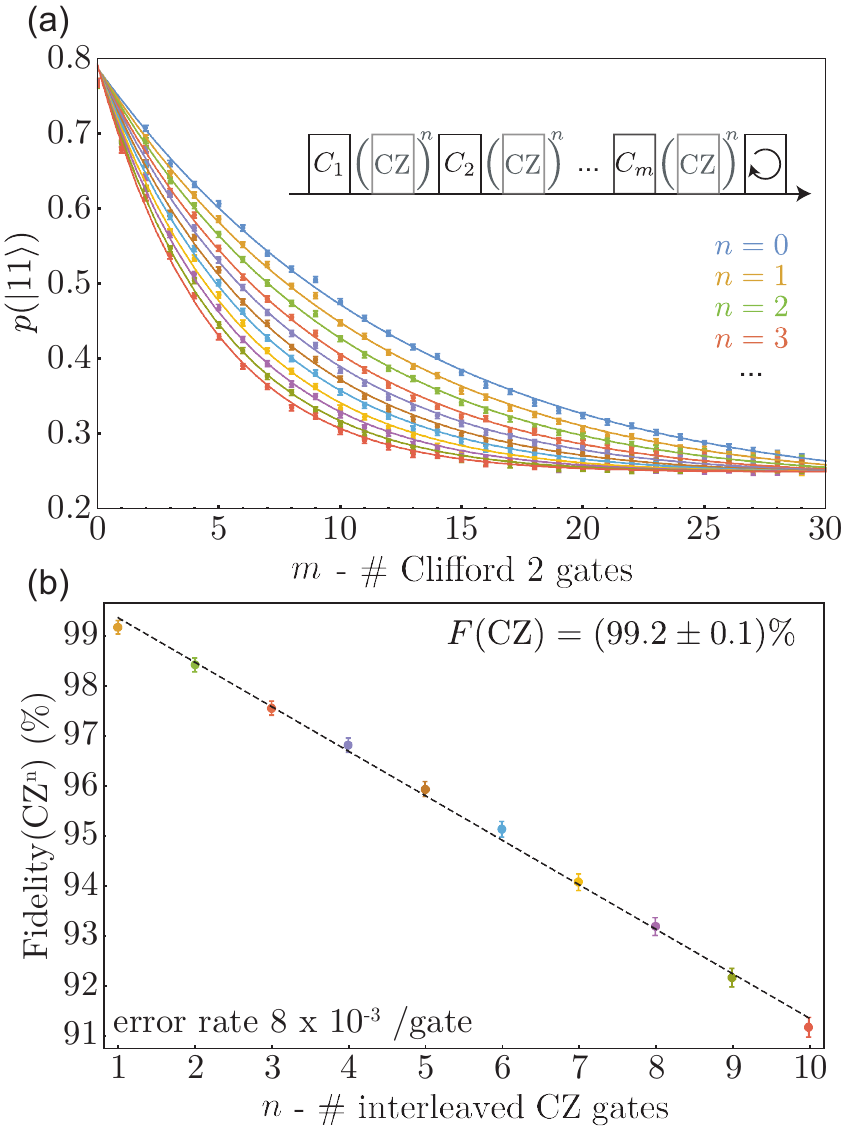}
    \caption{Interleaved randomized benchmarking (a) Average probability $p(|11 \rangle)$ as a function of sequence length. For each sequence length, we average over $100$ random sequences repeated $1500$ times. We insert a variable number $n=0,1, ...,10$ (encoded in the color) of CZ gates between the randomly chosen Clifford gates. (b) Fidelity of $\mathrm{CZ}^n$ versus $n$. The gate error grows linearly with $n$ indicating that the errors are incoherent. Error bars are calculated according to the procedure given in Appendix \ref{sec:errorbars}.}
    \label{fig:fig4}
\end{figure}

To reach the optimal gate fidelity, we perform a parameter search optimizing the sequence fidelity of fixed length interleaved randomized sequences \cite{Magesan2012a} using a method known as 'Optimized Randomized Benchmarking for Immediate Tune-up' (ORBIT) \cite{Kelly2014} with a Covariance Matrix Adaptation Evolution Strategy (CMA-ES) \cite{hansen2019pycma}. In practice, we first measure an interleaved randomized benchmarking curve [see Fig.~\ref{fig:fig3}(b)] before fixing the number of gates to $n = \mathrm{floor}[-1/\log(p)] = 5$ providing the optimal sensitivity. The survival probability $p(\left| 11 \right>)$ is maximized by adjusting the $6$ gate parameters [Fig.~\ref{fig:fig3}(a)]: amplitude, duration, width of the edges, frequency, and single-qubit rotation angles. Fig.~\ref{fig:fig3}(c-h) shows the stochastic evolution of gate parameters during $1700$ search steps leading to an improvement of gate fidelity from $0.967$ to $0.992$ for final gate duration of $30.0 \mathrm{(plateau)}+2 \times 15.8 \mathrm{(edges)} = 61.6 \mathrm{~ns}$. In our experiment, CMA evolution strategy leads systematically to the same (global) minimum contrary to non-stochastic algorithm such as Nelder-Mead method that gets more easily trapped in local minima.

Finally, we assess the quality of the optimized CZ gate by iterative interleaved randomized benchmarking \cite{Magesan2012a}. We obtain a reference fidelity $F_\mathrm{Clifford2} = (96.0\pm0.1)\%$. Interleaving the CZ gate [yellow curve in Fig.~\ref{fig:fig4}(a)] yields a gate fidelity of $F(\mathrm{CZ})= (99.2 \pm 0.1)\%$. This measurement is repeated for an increasing number of interleaved CZ gates [Fig.~\ref{fig:fig4}(b)] to separate the contribution of systematic coherent errors from decoherence effects \cite{Sheldon2015}. In our experiment, the total error of the CZ gate repeated $n$ times grows linearly with the number of gates [Fig~\ref{fig:fig4}(b)] demonstrating that the remaining error is incoherent and thus not correctable with coherent controls.

\section{\label{sec:discussion}Outlook}

Although the benchmarked fidelity of our CZ-gate is already high, the current experimental setup contains a number of imperfections, most of which can be readily eliminated in the next generation experiments. Let us start the discussion by briefly summarising these imperfections. 

Taking into account the Ramsey coherence times of transitions $|10 \rangle - |20 \rangle$ and $|11 \rangle - |21 \rangle$ (see Table \ref{table:T1T2}), one expects the gate error bounded by $0.7 \%$ (see Appendix \ref{sec:decoherenceeffect}). This estimate is compatible with the gate error $(0.8\pm 0.1) \%$ measured with randomized benchmarking. We believe that the limited coherence time of our higher-level transitions originates in the first order sensitivity to external flux noise caused by the sweet spot misalignment. This effect can be corrected by either improving magnetic field uniformity, likely through better magnetic shielding, or by using two independent coils. Importantly, fluxonium's $|1 \rangle - | 2 \rangle$ transition is very similar in term of decoherence mechanisms to a regular transmon transition \cite{Nguyen2018} and we therefore expect no fundamental obstacles in reaching coherence times around $50-100~\mu\textrm{s}$, on par with optimally fabricated flux-tunable transmons. In this case, the Hamiltonian simulations presented in Fig.~\ref{fig:fig2} shows that coherent gate errors in the low $ 10^{-4}$-range are possible for our device without sophisticated pulse optimization.

Coherence time in the computational subspace can exceed $500~\mu\textrm{s}$, given the measured energy relaxation times of $250-350~\mu\textrm{s}$ of states $|10\rangle$ and $|01\rangle$. The proof-of-principle demonstration of $T_2 \approx 500~\mu\textrm{s}$ coherence time was reported in a previous single-fluxonium experiment. Yet, here the qubit coherence time is reduced to $50-60~\mu\textrm{s}$, even at their flux sweet-spots. This is likely due to a residual average thermal photon population $n_\mathrm{th} \sim 5 \times 10^{-3}$ in the readout resonator. Coherence time of the gating transitions $|10\rangle-|20\rangle$ and $|11\rangle - |21\rangle$ will be limited eventually by thermal dephasing as well. This dephasing source is generic to all superconducting circuit experiments and it can be mitigated in the future with improved cryogenic filtering of the microwave measurement lines \cite{Wang2018,Yan2018}. After this technical task is accomplished, the qubits in our gate scheme would be considerably better than transmons at storing quantum information in between the two-qubit gate operations.

Turning to our microwave packaging choice, we used a single input port in a single 3D-box resonator to perform all the gates, for the sake of technical simplicity (see Appendix \ref{sec:setup}). In Appendix \ref{sec:singlequbitgates}, we characterize the reduction of addressability in the system and find that simultaneous single qubit gates have about a $\sim 3 \%$ error rate dominated by classical cross-talks. Using a dedicated driving port per qubit would enable selective addressing of each qubit and hence improve significantly the single qubit gates fidelity \cite{Gambetta2012}. More generally, individual addressing and readout is essential for scaling beyond the two-qubit experiments. We believe that scaling of our gate scheme can be done by simply moving to 2D-circuit designs, in complete analogy with processors based on capacitively-coupled transmon circuits \cite{Kelly2015,Otterbach2017,Kandala2017}. Indeed, our experiment does not benefit from the usually high quality factors of 3D-resonators, and hence it is compatible with a traditional 2D-circuit technology without conceptual modifications. The only explicit price of replacing fixed-frequency transmons by fluxoniums would be the requirement of either a highly homogeneous global magnetic field or a static flux-bias line per qubit.  

As a final remark, we stress that our CZ-gate does not require excessive parameter matching, unlike the majority of microwave-activated gates \cite{Chow2013a,Tripathi2019}. For example, the level diagram in Fig.~\ref{fig:fig1} has no fine-tuned transitions, and the described gate protocol would work for a large class of fluxonium spectra and interaction strengths. Essentially, one only need to make sure the states $|12\rangle$ and $|21\rangle$ are not too far from each other (around $300~\textrm{MHz}$ here) and that qubit frequencies are well resolved (about $70~\textrm{MHz}$ here). This property can mitigate the effects of fabrication imperfections and improve qubit connectivity in the large-scale quantum processors.

\section{\label{sec:conclusion}Conclusions}

In this paper, we demonstrated that low-frequency fluxonium qubits are not only good at storing quantum information, but they also allow for fast and high-fidelity logical operations with a minimal engineering overhead. Our implementation of the microwave-activated CZ gate can be decomposed into two ideas applicable to other quantum systems, each harnessing the strong anharmonicity of fluxoniums. First, the energy scale $\Delta$ limiting the gate speed comes from the repulsion of non-computational states (here the repulsion of states $|21\rangle$ and $|12\rangle$), and hence the gate time $\sim 1/\Delta$ can be made, in principle, shorter than the qubit's Larmor periods. Second, by synchronizing Rabi rotations of the only two relevant non-computational transitions (here $|10\rangle - |20\rangle$ and $|11\rangle - |21\rangle$), a conditional geometric phase is accumulated at the end of each Rabi period with zero leakage outside the computational manifold. Notably, the synchronization is possible with a single microwave pulse of proper amplitude, frequency, and duration, applied to either qubit. We adjusted the conditional phase to $\pi$ to get the CZ-gate, but the phase can be changed to any other value, resulting in the CPhase-gate.

We exhaustively characterized the CZ gate using both quantum process tomography and randomized benchmarking techniques. While the current combination of gate time, $61.6~\textrm{ns}$, and gate infidelity, $8\times 10^{-3}$, is already competitive with gates on transmons, our analysis indicates that the error can be reduced by an order of magnitude on improving the fabrication procedures, magnetic shielding, and line filtering. Given the compatibility of our capacitive coupling scheme with transmon-based scaling technology, we believe that all necessary demonstrations have been made to start exploring large-scale fluxonium-based quantum processors.

\begin{acknowledgments}
We thank Chen Wang, Benjamin Huard, and Ivan Pechenezhskiy for useful discussions and acknowledge the support from NSF PFC at JQI and ARO-LPS HiPS program. V.E.M. and M.G.V acknowledge the Faculty Research Award from Google and fruitful conversations with the members of the Google Quantum AI team.
\end{acknowledgments}

\appendix

\section{Device design and fabrication}

The device is fabricated by standard e-beam lithography of a resist bilayer (MAA/PMMA) deposited on a high-resistivity silicon substrate. Two layers of Aluminum ($20$ and $40 \mathrm{~nm}$) \cite{Nguyen2018} separated by a thin barrier of oxide are deposited by double angle evaporation. Each inductance is composed of 310 (qubit A) and 206 (qubit B) identical Josephson junction (in agreement with the extracted parameters $ E_{L,B}/E_{L,A}\simeq 206/310$). The loops are closed by a small Josephon junctions whose insulating barrier parameters determines $E_{J,A(B)}$. The antennas are designed to provide a charging energy $E_{C,A}, E_{C,B} \simeq 1 \mathrm{~GHz}$ required to bring the second excited state transitions in the range $4-5 \mathrm{~GHz}$. The asymmetry in the antenna design enables us to independently adjust the charging energy and the coupling to the cavity mode.

Table \ref{table:T1T2} gives the value of the energy relaxation and coherence times of the selected transitions of the system. All the lifetimes and coherence times are measured by energy relaxation, Ramsey and spin echo sequences. The reduced coherence time of the $|10 \rangle - |20 \rangle$ and $|11 \rangle - | 21 \rangle$ is limiting the gate fidelity (see main text).

\begin{center}
\begin{table}[h!]
\begin{tabular}{ | c | c | c | c | c | } \hline
  Transition & $\mathrm{freq ~(GHz)}$ & $T_1 ~\mathrm{(\mu s)}$ & $T_2^R ~\mathrm{(\mu s)}$ & $T_2^E ~\mathrm{(\mu s)}$ \\ \hline
  $|00 \rangle - |10 \rangle$ & 0.07233 & 347 & 5.6 & 31 \\ \hline
  $|00 \rangle - |01 \rangle$ & 0.13641 & 282 & 25.4 & 64\\ \hline
  $|10 \rangle - |20 \rangle$ & 5.1766 & 8.9 & 2.5 & 9.3 \\ \hline
  $|11 \rangle - |21 \rangle$ & 5.1986 & 6.1 & 1.7 & 4.3 \\ \hline
\end{tabular}
\caption{\label{table:T1T2} Frequencies and coherence times of transitions participating in the gate operation.
}
\end{table}
\end{center}

\section{Measurement setup and wiring \label{sec:setup}}

\begin{figure}
    \centering
    \includegraphics[width=\columnwidth]{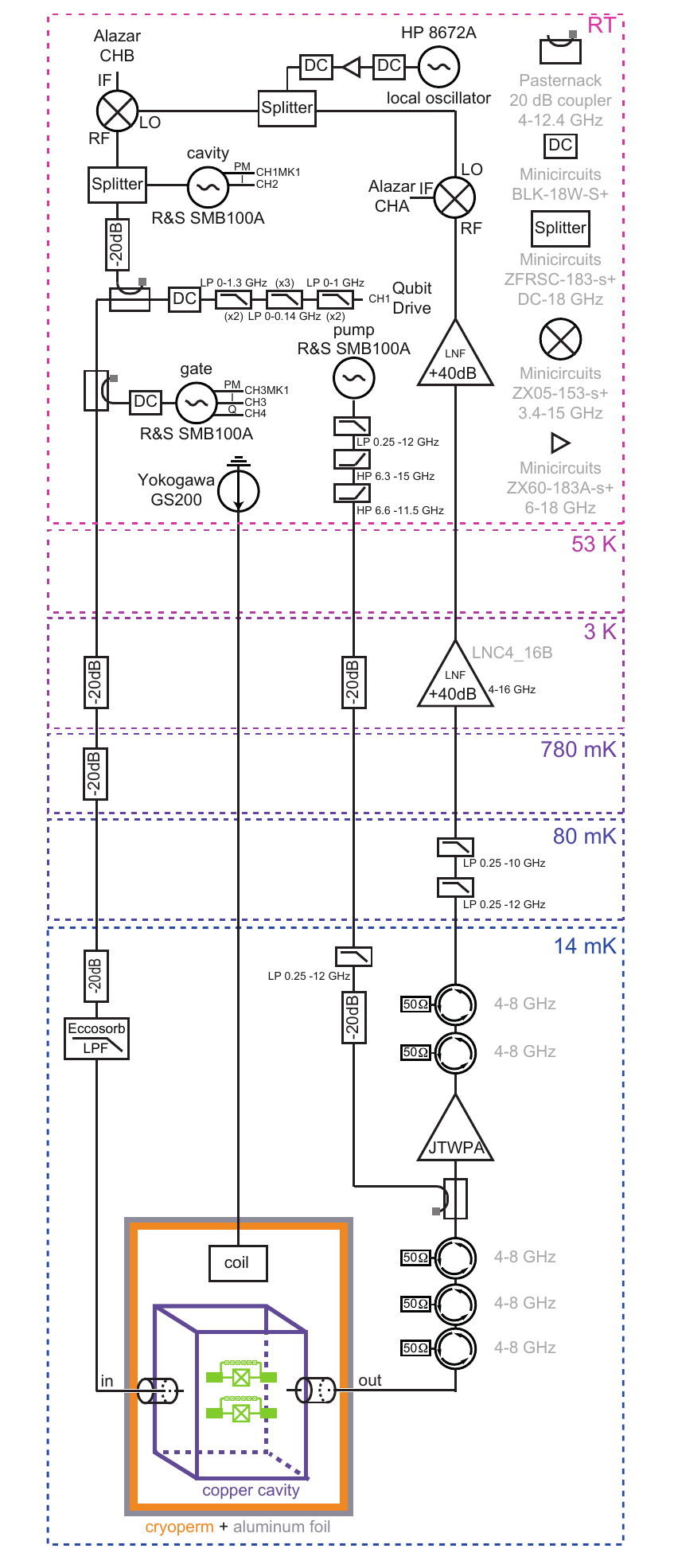}
    \caption{Schematics of the experimental setup. Single qubit pulses are generated directly at the qubit frequencies using one analog output port of a Tektronix\textregistered Arbitrary Waveform Generator AWG5014C (not represented). The signal is combined with CZ gate pulses and readout pulses before reaching the input port of the 3D cavity. The outgoing signal is amplified using a Traveling Wave Parametric Amplifier followed by cryogenic and room temperature amplifiers before down-conversion by a local oscillator, digitization by an Alazar\textregistered acquisition board (not represented), and numerical demodulation.}
    \label{fig:wiring}
\end{figure}

Qubit pulses are directly generated by an Arbitrary Waveform Generator (see Fig.~\ref{fig:wiring}) instead of modulating a radio frequency tone. Our ability to multiplex and synthesize qubit pulses with a single digital-to-analog converter significantly reduces the hardware cost of the experiment. We use the internal mixing and pulse modulation capabilities of two Rhode and Schwarz\textregistered SMB100A sources to generate the readout pulse and the two qubit operations. The output signal is amplified by a Josephson Traveling Wave Parametric Amplifier (JTWPA) provided by Lincoln Labs \cite{Macklin2015} followed by commercial amplifiers before heterodyne detection.

\section{Initialization and readout \label{sec:initializationandreadout}}

\subsection{Single-shot readout \label{sec:singleshotreadout}}

\begin{figure}
    \centering
    \includegraphics[width=\columnwidth]{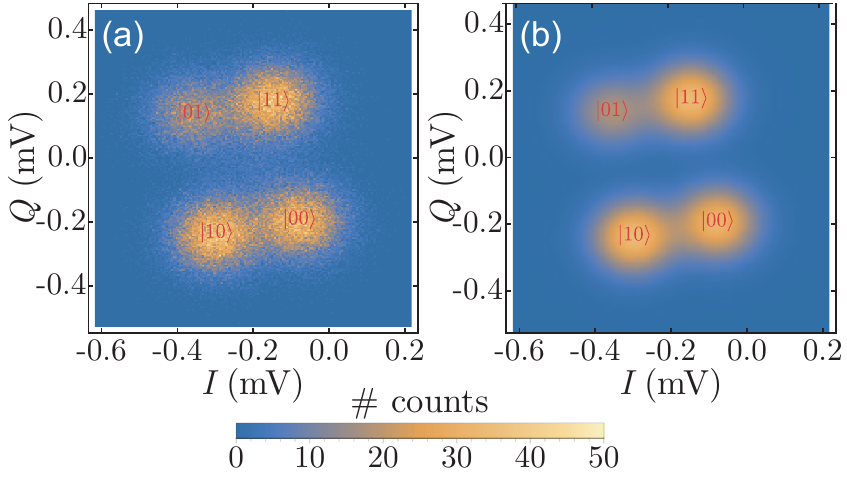}
    \caption{Experimental (a) and fitted (b) single-shot histograms of the readout signal when the two qubits are in thermal equilibrium. Histograms in (a) are calculated from $1.5 \times 10^5$ experimental realizations with a $10 \mathrm{~ \mu s}$-long readout pulse The fitting function is the sum of four gaussian distributions. This measurement is used to calibrate the position of the center of the gaussian distributions.}
    \label{fig:singleshot}
\end{figure}

Both fluxoniums are strongly coupled to a 3D cavity. We perform a joint readout \cite{Filipp2009,Dicarlo2009} of the qubit states. In the dispersive regime \cite{Zhu2013}, the readout operator can be written
\begin{equation}
\hat{M} = \beta_{IZ} \mathbb{I} \otimes \sigma_Z+\beta_{ZI} \sigma_Z \otimes \mathbb{I}+\beta_{ZZ} \sigma_Z \otimes \sigma_Z
\end{equation} where $\beta_{ij}$ are complex coefficients and $\sigma_i$ are Pauli matrices. In the single-shot limit, the integrated heterodyne signal distribution can be modeled as the sum of four Gaussian distributions associated to the computational states $| 00 \rangle$, $| 10 \rangle$, $| 01 \rangle$, $| 11 \rangle$ (see Fig.~\ref{fig:singleshot}).

\begin{figure}
    \centering
    \includegraphics[width=\columnwidth]{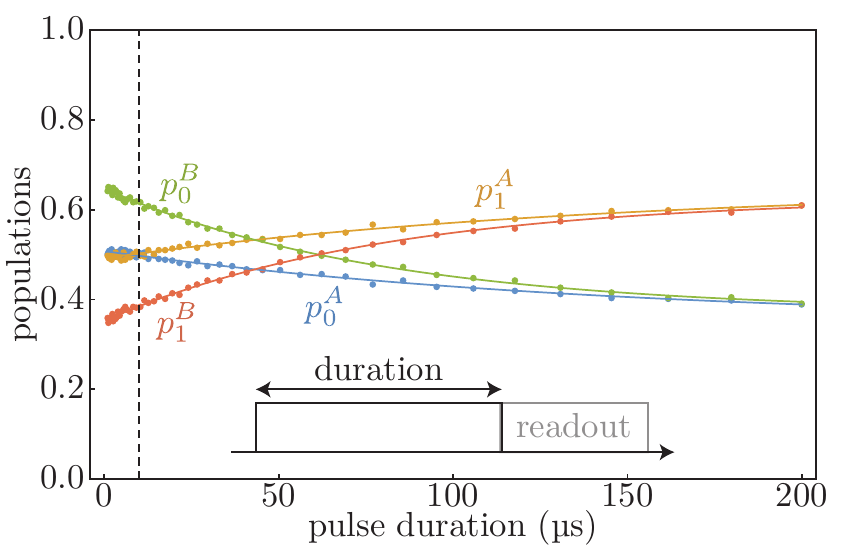}
    \caption{Evolution of the qubit populations after a cavity pulse as a function of the pulse duration. The solid lines correspond to the solution of a rate equation Eq.~\ref{eq:rate} for each qubit with the transition rate $\Gamma_\uparrow^A = 4.4 \pm 0.1 \mathrm{~kHz}$, $\Gamma_\downarrow^A = 2.4 \pm 0.1 \mathrm{~kHz}$ and $\Gamma_\uparrow^B = 7.83 \pm 0.05 \mathrm{~kHz}$, $\Gamma_\downarrow^B = 4.64 \pm 0.05 \mathrm{~kHz}$. The dashed line indicates the duration of the readout pulse. Note that the impact of the readout pulse in this figure is already corrected by the method described in section \ref{sec:readoutcrosstalk}.}
    \label{fig:nonQNDness}
\end{figure}

Surprisingly, we observe that the populations (Fig.~\ref{fig:singleshot}) extracted from fitting the readout distribution by four Gaussians are affected by the readout amplitude and duration indicating that our readout scheme is not Quantum Non-Demolition (QND) \cite{Boissonneault2009,Slichter2012,Sank2016}. Fig.~\ref{fig:nonQNDness} represents the evolution of the population with the duration of a cavity pulse preceding the readout pulse. 

Although recent studies demonstrated the QND readout of granular aluminum fluxonium circuits with hundreds of photons \cite{Gusenkova2020}, the non-QNDness of the readout of junction based fluxoniums was observed in the past either directly \cite{Vool2014} or in the form of a qubit lifetime suppression by readout photons \cite{Manucharyan2012a} even for average stationary photon number of the order of unity. These mechanisms are not investigated here and will be the subject of further studies.

\subsection{Readout cross-talks compensation \label{sec:readoutcrosstalk}}

We adopt an empirical readout cross-talk compensation introduced in Ref.~\cite{Dewes2012}. This approach was used to compensate incorrect state mapping during a bifurcation readout but we believe that it is relevant in our case. First, we compensate for an incorrect mapping $\left| 0 \right> \rightarrow \left| 1 \right>$ or $\left| 1 \right> \rightarrow \left| 0 \right>$ of qubit $\alpha$ by correcting qubit population of states $|ij \rangle$ with a tensorial products of two $2 \times 2$ matrices $C_A \otimes C_B$ where $C_\alpha = \begin{pmatrix}
a_\alpha & 1-b_\alpha \\
1-a_\alpha & b_\alpha
\end{pmatrix}$ where $a_{A(B)} = 0.98 (0.96)$ and $b_{A(B)} = 0.96(0.87)$.

\begin{figure}
    \centering
    \includegraphics[width=\columnwidth]{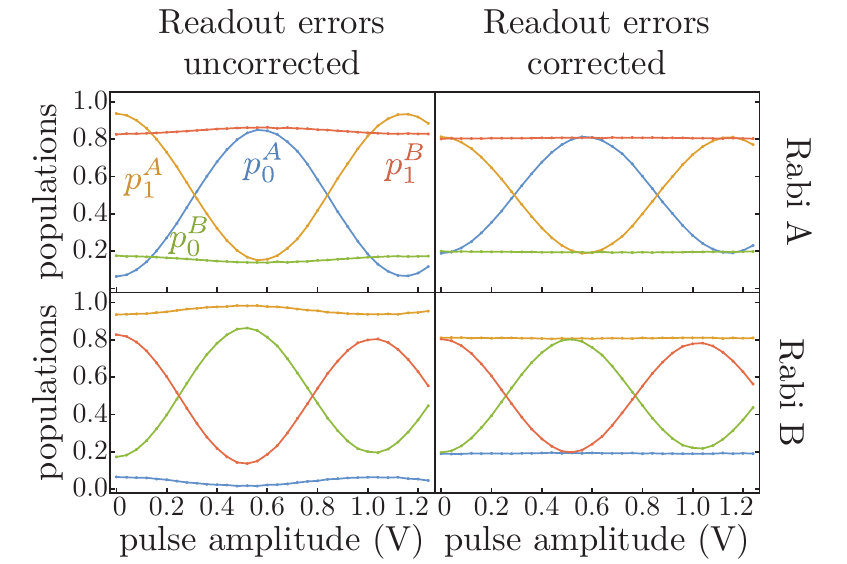}
    \caption{Rabi oscillations of qubit A (first row) or qubit B (second row) without (left column) and with (right column) readout error correction. After the calibration, only the target qubit displays oscillations.
    }
    \label{fig:readoutxtalks}
\end{figure}

In addition, we use a pure cross-talk matrix which takes into account the possibility to swap excitations between the qubits during the readout process. The total readout correction applied to the qubit populations reads
\begin{equation}
    \begin{pmatrix}
    p_{00}'\\
    p_{10}'\\
    p_{01}'\\
    p_{11}'
    \end{pmatrix} 
    = 
    \begin{pmatrix}
    1&0&0&0 \\
    0&1-b&c&0\\
    0&b&1-c&0\\
    0&0&0&1
    \end{pmatrix}
    (C_A
    \otimes
    C_B)
    \begin{pmatrix}
    p_{00}\\
    p_{10}\\
    p_{01}\\
    p_{11}
    \end{pmatrix}
    \label{eq:readoutcorrection}
\end{equation} where $p_{ij}'$ are the corrected qubit populations, $b= 7\%$ and $c = 3.5 \%$. The impact of this calibration is best exemplified when performing Rabi experiments. After the calibration (Fig.~\ref{fig:readoutxtalks} right column), we observe that the oscillations are centered around $0.5$ (center of the Bloch sphere) and only the targeted qubit displays oscillations. Finally, we remind the reader that the two-qubit gate fidelities quoted in the main text are not affected by this procedure since randomized benchmarking is not sensitive to readout errors \cite{Magesan2012}.

\subsection{Initialization \label{sec:initialization}}

We describe how we turned the qubit transitions induced by cavity photons observed in Fig.~\ref{fig:nonQNDness} to our advantage to initialize the system. The rate of these transitions is known to generally increase with the number of circulating photons in the cavity \cite{Manucharyan2012a,Gusenkova2020}. We thus expect to be able to induce incoherent transition rates between the qubit states.

\begin{figure}
\centering
\includegraphics[width=\columnwidth]{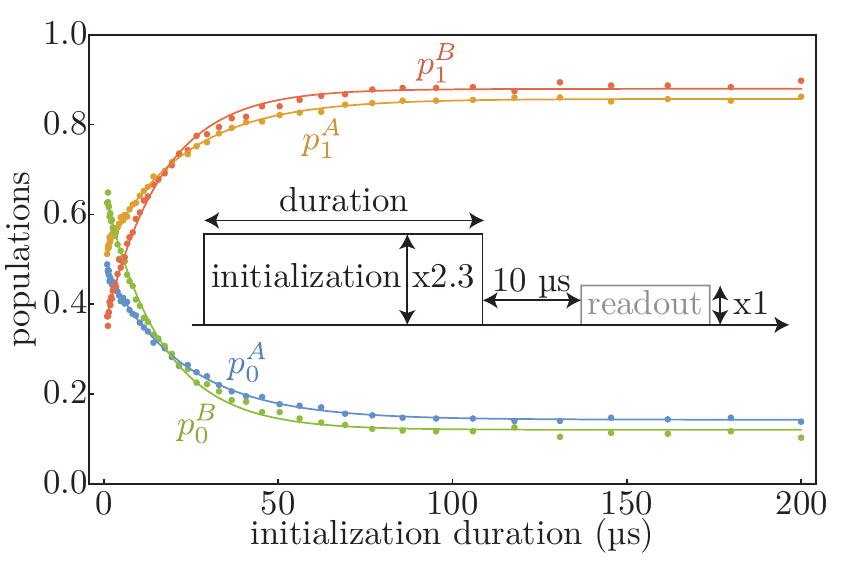}
\caption{Initialization of the two fluxoniums. We apply a large ($2.3$ times the readout amplitude) drive on the cavity in order to reduce the entropy of the system. Populations are extracted by the method described in section \ref{sec:initialization}. The solid lines correspond to a fit to rate equations for each qubit with following transition rates $\Gamma_\uparrow^A = 38.6 \pm 0.6 \mathrm{~kHz}$, $\Gamma_\downarrow^A = 6.4 \pm 0.6 \mathrm{~kHz}$ and $\Gamma_\uparrow^B = 53.8 \pm 0.3 \mathrm{~kHz}$, $\Gamma_\downarrow^B = 7.4 \pm 0.3 \mathrm{~kHz}$. In all the measurements shown in this paper, we apply a $200 \mathrm{~\mu s}$ initialization pulse that leaves the qubits in mixed states with excitation probability $p_1^A = 0.86 $ and $p_1^B = 0.88$ that correspond to reversed thermal states with temperatures $T_A = 3.7 \mathrm{~mK}$ and $T_B = 5.9 \mathrm{~mK}$. The entropy extracted by the initialization pulse is $\Delta S / \ln(2) = 0.81 \mathrm{~bits}$.}
\label{fig:initialization}
\end{figure}

We model the photon-induced relaxation using the incoherent rate equations
\begin{equation}
\begin{aligned}
&\frac{d p_0^\alpha}{d t} = -\Gamma_\uparrow^\alpha p_0^\alpha + \Gamma_\downarrow^\alpha p_1^\alpha \\
&\frac{d p_1^\alpha}{d t} = \Gamma_\uparrow^\alpha p_0^\alpha - \Gamma_\downarrow^\alpha p_1^\alpha 
\end{aligned}
\label{eq:rate}
\end{equation}where the total excitation and de-excitation rates are $\Gamma_\uparrow^A = 38.6 \pm 0.6 \mathrm{~kHz}$, $\Gamma_\downarrow^A = 6.4 \pm 0.6 \mathrm{~kHz}$ and $\Gamma_\uparrow^B = 53.8 \pm 0.3 \mathrm{~kHz}$, $\Gamma_\downarrow^B = 7.4 \pm 0.3 \mathrm{~kHz}$. Notably, increasing the number of circulating photons by a factor $2.3^2 \sim 5$ (from Fig.~\ref{fig:nonQNDness} to Fig.~\ref{fig:initialization}) increases the energy relaxation rates $\Gamma_\downarrow^A+\Gamma_\uparrow^A$ and $\Gamma_\downarrow^B+\Gamma_\uparrow^B$ by about a factor $\sim 5$ mainly by increasing the excitation rates. The amplitude of the initialization pulse was chosen to maximize the steady state purity while keeping the qubit populations in the computational space. This procedure leads to the preparation of the qubit states in a statistical mixture with the excited state probabilities $p_1^A = 0.86 $ and $p_1^B = 0.88$ (see Fig.~\ref{fig:initialization}).

\section{Single qubit gate fidelities \label{sec:singlequbitgates}}

\begin{figure}
    \centering
    \includegraphics[width=\columnwidth]{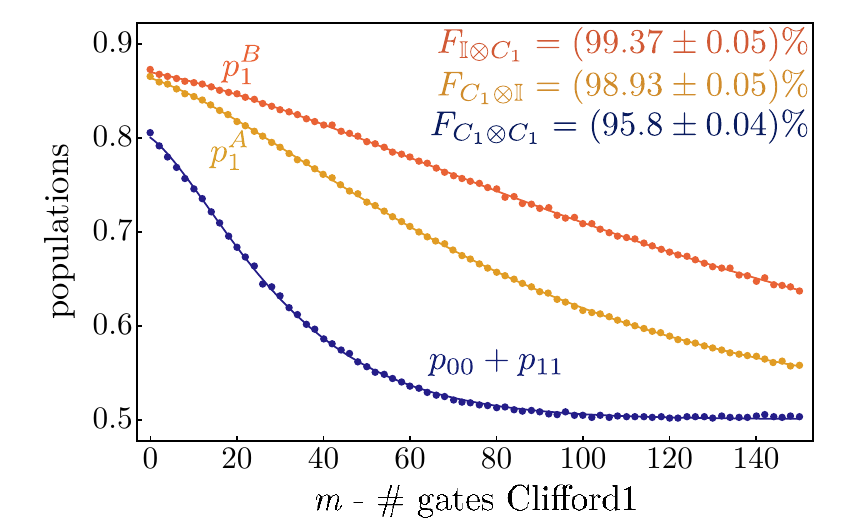}
    \caption{Randomized benchmarking for single qubit gates on qubit A, qubit B and both. The solid lines are fit by a first order model \cite{Magesan2012} $a+b p^m + c (m-1) p^{m-2}$ where $p$ is the depolarizing parameter used to calculate the fidelity $F=1-\frac{d-1}{d}(1-p)$ where $d$ is the dimension of the Hilbert space. Error bars on the fidelity are calculated according to the procedure described in Appendix \ref{sec:errorbars}.}
    \label{fig:singlequbitgates}
\end{figure}

Single qubit gate fidelities are calibrated by single qubit randomized benchmarking (see Fig.~\ref{fig:singlequbitgates}). Qubit A (B) gates are generated using Gaussian edge pulses with a total duration of $150 \mathrm{~ns}$ ($75~\mathrm{~ns}$). Reducing the gate duration of single qubit gates deteriorates the simultaneous randomized benchmarking fidelity because of cross-talks.

The addressability of each qubit can be characterized by the numbers in table \ref{table:addressability} using the definitions in Ref.~\cite{Gambetta2012}. The reduction of addressability in our system is related to the fact that our control field aimed at one target qubit is also influencing the other qubit (classical cross-talk) due to our choice of a sample design with a single input port.

\begin{center}
\begin{table}[!ht]
\begin{tabular}{ | c | c | c | } \hline
   parameter & Twirl group & experimental value \\ \hline
   $r_A$ & $\mathrm{Clifford1}\otimes \mathbb{I}$ & $6.3 \times 10^{-3}$ \\ \hline
   $r_B$ & $\mathbb{I} \otimes \mathrm{Clifford1}$& $1.07 \times 10^{-2}$ \\ \hline
   $r_{A|B}$ & $\mathrm{Clifford1} \otimes \mathrm{Clifford1}$ & $5.3 \times 10^{-3}$ \\ \hline
   $r_{B|A}$ & $\mathrm{Clifford1} \otimes \mathrm{Clifford1}$ & $1.4 \times 10^{-2}$ \\ \hline
   $\delta r_{A|B}$ & & $8.1 \times 10^{-3}$ \\ \hline
   $\delta r_{B|A}$ & & $5.4 \times 10^{-3}$ \\ \hline
   $\delta p$ & & $2.2 \times 10^{-2}$\\ \hline
\end{tabular}
\caption{\label{table:addressability} Table of addressability metrics using the definition of Ref.~\cite{Gambetta2012}. The error rates $r_\alpha$ are extracted from the independent randomized benchmarking experiments of Fig.~\ref{fig:singlequbitgates}. The error rate $r_{\alpha|\beta}$ are extracted by looking at the depolarization of qubit $\alpha$ during a simultaneous randomized benchmarking experiment. The error metric of error on $\alpha$ due to the unwanted control of $\beta$ is given by  $\delta r_{\alpha|\beta} = |r_\alpha-r_{\alpha|\beta}|$. The correlations in errors are flagged by $\delta p = p_{AB} - p_{A|B} p_{B|A}$ where $p_{AB}$ is obtained from fitting the decay of qubit-qubit correlations $p_{00} + p_{11}$.}
\end{table}
\end{center}

The use of a first order model \cite{Magesan2012} to fit the randomized benchmarking decay can be justified by the presence of a time-dependent gate dependent noise (flux noise) \cite{Wallman2018}. Table \ref{table:gatefidelities} summarizes the fidelity of various gates used to generate the Clifford group.

\begin{center}
\begin{table}[h!]
\begin{tabular}{ | c | c | } \hline
  gate & fidelity ($\%$) \\ \hline
  $X_{\pi/2}-\mathbb{I}$ & $97.9 \pm 0.1$\\ 
  $Y_{\pi/2}-\mathbb{I}$ & $97.9 \pm 0.1$\\ 
  $X_{\pi}-\mathbb{I}$ & $97.6 \pm 0.1$\\ 
  $Y_{\pi}-\mathbb{I}$ & $97.6 \pm 0.09$\\ 
  $\mathbb{I}-X_{\pi/2}$ & $98.95 \pm 0.08$\\ 
  $\mathbb{I}-Y_{\pi/2}$ & $98.96 \pm 0.08$\\ 
  $\mathbb{I}-X_{\pi}$ & $99.20 \pm 0.08$ \\ 
  $\mathbb{I}-Y_{\pi}$ & $99.20 \pm 0.08$\\ 
  $X_{\pi/2}-X_{\pi/2}$ & $97.8 \pm 0.1$\\ 
  $X_{\pi/2}-Y_{\pi/2}$ & $97.8 \pm 0.1$\\ 
  $Y_{\pi/2}-X_{\pi/2}$ & $97.8 \pm 0.1$\\ 
  $Y_{\pi/2}-Y_{\pi/2}$ & $97.8 \pm 0.1$\\ 
  $X_{\pi}-X_{\pi}$ & $93.9 \pm 0.2$\\ 
  $X_{\pi}-Y_{\pi}$ & $94.0 \pm 0.2$\\ 
  $Y_{\pi}-X_{\pi}$ & $93.8 \pm 0.2$\\ 
  $Y_{\pi}-Y_{\pi}$ & $94.2 \pm 0.2$ \\ \hline
\end{tabular}
\caption{\label{table:gatefidelities} Fidelities of various Clifford gates obtained by two-qubit interleaved randomized benchmarking.}
\end{table}
\end{center}

\section{Error bars \label{sec:errorbars}}

The error bars displayed in Fig.~\ref{fig:fig4} are the standard deviations on the population obtained from the population extracted from the method described in Appendix \ref{sec:initializationandreadout}. All the curves in Fig.~\ref{fig:fig4}(a) are fitted simultaneously to a first order model $a + b p(n)^m + c (m-1) p(n)^{m-2}$ where $a,b,c$ are fitting parameters used to absorb state preparation are measurement (SPAM) errors, $m$ is the number of Clifford gates and $n$ is the number of interleaved CZ gates. The fidelity of a $n$ iteration of a CZ gate is given by $F(\mathrm{CZ}^n) = 1-\frac{d-1}{d} \left(1-\frac{p(n)}{p(n=0)} \right)$ with $d=4$.

The error on the gate fidelity is then estimated by the propagation of error formula
\begin{equation}
\Delta F(\mathrm{CZ}^n) = \frac{d-1}{d} \sqrt{ \left(\frac{\Delta p(n)}{p(n=0)}\right)^2+\left( \frac{p(n) \Delta p(n=0)}{p(n=0)^2} \right)^2}
\end{equation} to obtain the error bars in Fig.~\ref{fig:fig4}(b). Similar procedures are used to obtain all the error bars given in the paper.

\section{Theory and simulations of gate errors\label{sec:driveparameters}}
\subsection{Unitary errors}

To study the unitary dynamics during the gate operation, we solve the Schr\"odinger equation numerically for the time-dependent Hamiltonian given by Eqs.~(\ref{eq:Hamiltonian}) and (\ref{eq:drive}) with the additional account for a Gaussian flat-topped pulse in the drive term (\ref{eq:drive}). Thus, we multiply the drive term by a pulse-shaping function with the rising edge  given by
\begin{equation}
f(t) \propto \exp\left[-(t-t_{\rm width})^2 / 2\sigma^2\right] - \exp\left[-t_{\rm width}^2 / 2\sigma^2\right]
\end{equation}
at $0<t < t_{\rm width}$,
where $\sigma = t_{\rm width}/\sqrt{2\pi}$. The lowering edge of the pulse is given by a symmetric expression at $t_{\rm width} + t_{\rm plateau} < t < t_{\rm gate}$, where $t_{\rm plateau}$ is the duration of the flat part and $t_{\rm gate} = 2t_{\rm width} + t_{\rm plateau}$ is the total gate duration. In all the simulations, we keep $t_{\rm width} = 15$ ns and vary $t_{\rm plateau}$ in order to vary $t_{\rm gate}$. To match the measured ratio between resonance Rabi frequencies $\Omega_{10-20}$ and $\Omega_{11-21}$, we choose $\epsilon_A / \epsilon_B = 0.9$ in the drive term~\eqref{eq:drive}.

\begin{figure}
    \centering
        \includegraphics[width=\columnwidth]{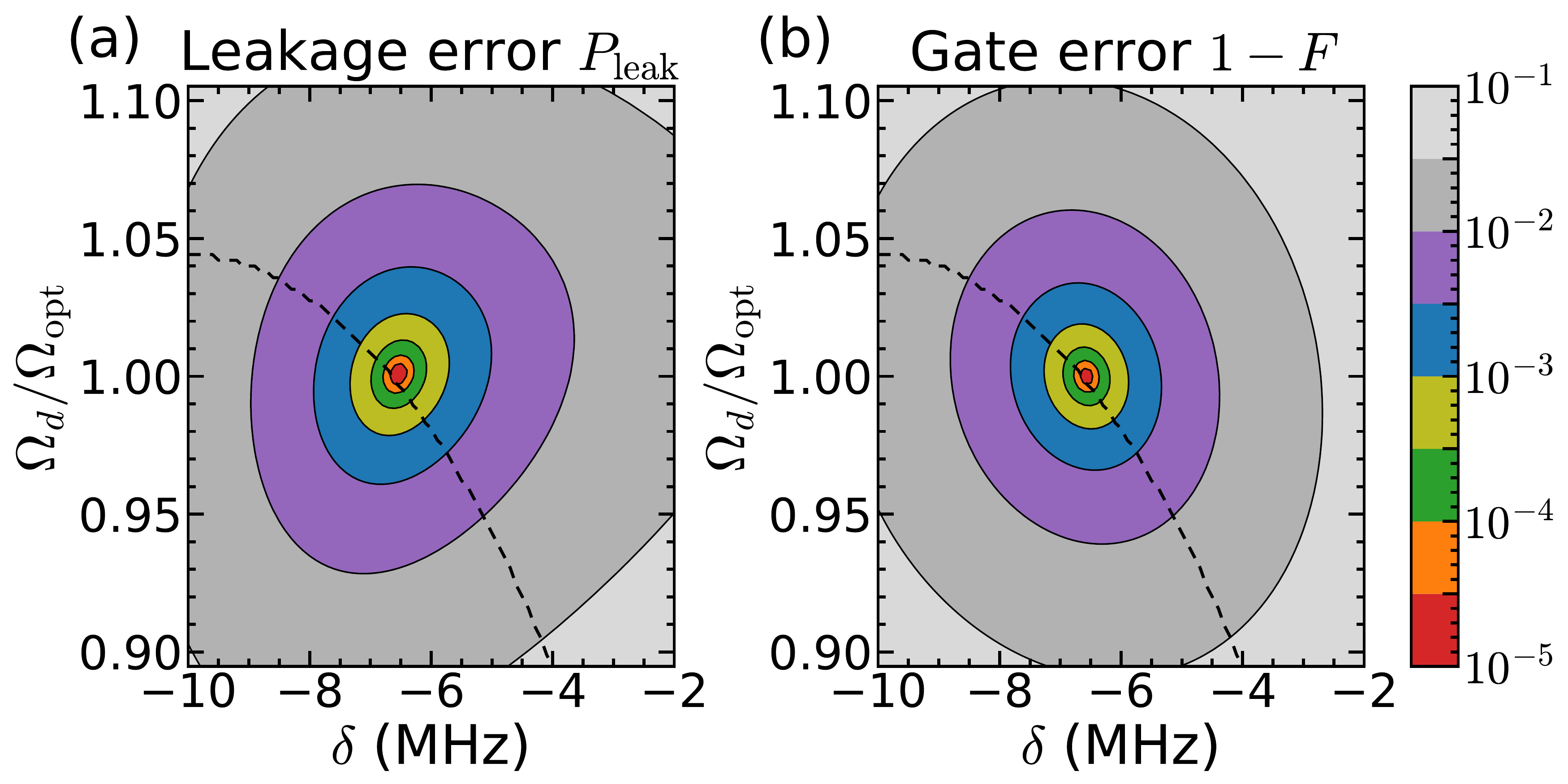}
    \caption{Simulated leakage error (a) and gate error (b) for the total gate duration of $t_{\rm gate}=61$ ns vs drive-frequency detuning $\delta$ and drive amplitude $\Omega_d$ in units of its optimal value $\Omega_{\rm opt}$. Dashed lines show parameter values satisfying the correct phase accumulation $\Delta\varphi = \pi$.}
    \label{fig:theory_2dcolor}
\end{figure}

We find the evolution operator $\hat{U}$ by calculating the evolution of four basis computational states, projecting the result into the computational subspace, and performing single-qubit Z rotations as described in Ref.~\cite{Nesterov2018}, which ensures that the only diagonal matrix element that has a nonzero phase is $\langle 11|\hat{U}|11\rangle$. We then calculate the average gate fidelity as~\cite{Pedersen2007}
\begin{equation}\label{eq:fidelity_def}
    F = \frac{{\rm Tr}\left(\hat{U}^\dagger \hat{U}\right) + \left|{\rm Tr}\left(\hat{U}_{\rm CZ}^\dagger \hat{U}\right)\right|^2}{20}\,,
\end{equation}
where $\hat{U}_{\rm CZ} = {\rm diag}(1, 1, 1, -1)$ is the diagonal operator for the ideal CZ gate. This expression corresponds to the randomized benchmarking fidelity \cite{Chow2009}. When the real gate operator is $\hat{U} = {\rm diag}(1, 1, 1, e^{i\Delta\varphi})$, we find that $1-F = (3/10) ( 1 +\cos \Delta\varphi)$. For a generic gate operator, we define $\Delta\varphi = {\rm arg} \langle 11|\hat{U}|11\rangle$ and use the last expression to calculate the phase error, which is shown by dots in Fig.~\ref{fig:fig2}(c). Once projected into the computational subspace, the operator $\hat{U}$ in Eq.~(\ref{eq:fidelity_def}) is generally nonunitary, which describes coherent leakage to noncomputational levels. The average  probability of such leakage is given by
\begin{equation}\label{eq:leakage_error}
    P_{\rm leak} = {1 - \frac 14{\rm Tr} \left(\hat{U}^\dagger \hat{U}\right)}\,.
\end{equation}
We show this leakage error in Fig.~\ref{fig:theory_2dcolor}(a) as well as by the dashed line in Fig.~\ref{fig:fig2}(c). The total gate error calculated using Eq.~(\ref{eq:fidelity_def}) is shown in Fig.~\ref{fig:theory_2dcolor}(b) and by the solid line in Fig.~\ref{fig:fig2}(c). Results shown in these figures illustrate that we can achieve coherent gate and leakage errors below $10^{-4}$ with the current amplitude and frequency stability of the experiment.

\subsection{Decoherence effects \label{sec:decoherenceeffect}}

We estimate incoherent errors resulting from driving the  $|10 \rangle - | 20 \rangle$ and $| 11 \rangle - | 21 \rangle$ transitions as follows. We add the errors coming from cycling each transition independently. We first estimate the state fidelity for an arbitrary initial state when driving a full rotation on one of the transitions and average this fidelity over a set of 16 independent initial states before summing the two contributions. We solve the master equation by treating incoherent processes as a perturbation and write the density matrix as an expansion $\rho = \rho^{(0)}+\rho^{(1)}+ ...$. For the initial state $|\psi(0) \rangle = |\psi_\perp \rangle +  c |11\rangle $, where $|\psi_\perp \rangle$ describes the amplitudes of the remaining computational basis states, we consider the Hilbert space with only three levels $| \psi_\perp \rangle$, $|11\rangle$, and $|21\rangle$. Here, $\langle \psi_\perp |\psi_\perp \rangle = 1-|c|^2$. We consider the drive Hamiltonian on the $|11 \rangle - |21 \rangle$ transition in the interaction picture and under the rotating wave approximation
\begin{equation}\label{eq:Hamitonian_incoherent_theory}
\begin{split}
    \frac{\hat{H}_\Omega}{h} & = \frac{\Omega}{2}\left(|11\rangle \langle 21| + |21\rangle \langle 11|\right) \\
   & =  \frac{\Omega}{2}|+\rangle \langle +| -  \frac{\Omega}{2} |-\rangle \langle -| \,, 
\end{split}
\end{equation}
where we have introduced the basis $|\pm\rangle = (|11\rangle \pm |21\rangle / \sqrt{2}$ that diagonalizes $\hat{H}_\Omega$. The unitary evolution under this Hamiltonian results in
\begin{multline}
    |\psi(t)\rangle = | \psi_\perp \rangle + c\left[\cos\left({\pi\Omega t}\right) |11\rangle - i \sin\left({\pi\Omega t}\right) |21\rangle \right]\\
    = | \psi_\perp \rangle + c\left[\frac{ e^{-i\pi\Omega t}}{\sqrt{2}}|+\rangle +  \frac{ e^{i\pi\Omega t }}{\sqrt{2}}|-\rangle\right]\,,
\end{multline}
which gives us the zeroth-order approximation to the density matrix $\rho^{(0)}(t) = |\psi(t)\rangle \langle \psi(t)|$. We find the first-order correction $\rho^{(1)}(t)$ from the iterative master equation
\begin{equation}
    \frac{d\rho^{(1)}}{dt} = -\frac{i}{\hbar}\left[\hat{H}_\Omega,\rho^{(1)}(t) \right] + \mathcal{L}\rho^{(0)}(t)\,,
\end{equation}
where the Lindblad super-operator $\mathcal{L}\rho = \sum_k \left[\hat{L}_k \rho\hat{L}_k^\dagger - (1/2)(\hat{L}_k^\dagger\hat{L}_k\rho  + \rho \hat{L}_k^\dagger\hat{L}_k ) \right]$ describes nonunitary processes. We use the following collapse operators
\begin{subequations}
\begin{align}
\label{eq:theory_L_r}
 \hat{L}_1 &= \sqrt{\Gamma_1} |11\rangle \langle 21| 
\,, \\
 \label{eq:theory_L_phi}
 \hat{L}_\varphi &= \sqrt{2\Gamma_\varphi} |21\rangle \langle 21| 
\,.
\end{align}
\end{subequations}
The first operator describes relaxation of the $|11\rangle - |21\rangle$ transition with a rate $\Gamma_1=1/T_1(11 - 21)$, and the second operator describes pure dephasing with a rate $\Gamma_\varphi = 1/T^R_2(11 - 21) - 1/2 T_1(11-21)$ (see Table \ref{table:T1T2}).

This gives the matrix element of the correction in the basis diagonalizing $\hat{H}_\Omega$
\begin{equation}
    \left\langle m \left|\rho^{(1)}(t) \right| n\right\rangle = \int \limits_0^{t} \left\langle m\left| \mathcal{L}\rho^{(0)}(t) \right| n\right\rangle e^{-i\nu_{mn}(t-t')} dt'\,,
\end{equation}
where $h\nu_{mn} = E_m-E_n$ is the difference between eigenvalues of $\hat{H}_\Omega$ corresponding to $|m\rangle$ and $|n\rangle$. These eigenvalues belong to the set $\{\pm h\Omega/2, 0$\}.

 Using these operators and equations above, we find the following expression for the state-preparation error ($t_{\rm gate} = 1/\Omega$):
\begin{multline}\label{eq:state_fidelity}
    1 - F_\psi = -{\rm Tr}\left[\rho^{(0)}(t_{\rm gate})\rho^{(1)}(t_{\rm gate})\right] \\
    = \frac{(\Gamma_1 + 2\Gamma_\varphi)t_{\rm gate}}{2}|c|^2 ( 1- |c|^2) + \frac{(3\Gamma_1 + 2\Gamma_\varphi)t_{\rm gate}}{8} |c|^4\,.
\end{multline}

To find the gate error, we average Eq.~(\ref{eq:state_fidelity}) over 16 initial two-qubit states generated from single-qubit states $|0\rangle$, $|1\rangle$, $(|0\rangle + |1\rangle)/\sqrt{2}$, and $(|0\rangle + i|1\rangle)/\sqrt{2}$. This gives the average values $\langle |c|^2\rangle = 1/4$ and $\langle |c|^4\rangle = 9/64$, which results in the gate error
\begin{equation}
    1 - F = \frac{(55\Gamma_1 + 74\Gamma_\varphi)t_{\rm gate}}{512} = \frac{9}{256}\frac{t_{\rm gate}}{T_1} + \frac{37}{256}\frac{t_{\rm gate}}{T_2}\,,
\end{equation}
where we have used the relations $\Gamma_1 = 1/T_1$ and $\Gamma_\varphi = 1/T_2 - 1/2T_1$. The upper bound for the contribution coming from the $|10\rangle - |20\rangle$ has the same form.

Comparing the values of $T_1$ and Ramsey $T_2$ in Table~\ref{table:T1T2}, we find that the main contribution is coming from $T_2$, so $1-F \approx (37/256) t_{\rm gate}/T_2 \approx 0.15 t_{\rm gate} / T_2$. Then, for $t_{\rm gate} = 1/\Delta = 45.5$ ns
(our analysis is valid for a square pulse), we find that $1-F \approx 0.15  t_{\rm gate}(1 / T_2^R(10 -20) + 1 / T_2^R(11 -21) \approx 0.67\%$.

As a sanity check, we have also integrated numerically the master equation in the six-level Hilbert space that consists of the computational subspace and two upper levels $|20\rangle$ and $|21\rangle$. We considered the Hamiltonian generalized from Eq.~(\ref{eq:Hamitonian_incoherent_theory}) to include the drive of the $|10\rangle - |20\rangle$ transition and to account for detunings between the drive frequency and two transition frequencies as described in Sec.~\ref{sec:concept}.
Using collapse operators given in Eqs.~(\ref{eq:theory_L_r}) and (\ref{eq:theory_L_phi}) and similar operators for the $|10 \rangle - |20 \rangle$ transition, we find the incoherent error to be $0.64\%$, which is close to the analytic estimate of $0.67\%$. 

\bibliography{references}

\end{document}